\documentclass[aps,prl,superscriptaddress,twocolumn,showpacs,amsmath,amssymb,floatfix]{revtex4}

\usepackage{graphicx}
\usepackage{dcolumn}
\usepackage{bm}
\begin{document}
\title{Direct observation of the coexistence of pseudogap and superconducting \\ quasi-particles in Bi2212 by time-resolved optical spectroscopy}
\author{Y. H. Liu$^{*}$}

\affiliation{Department of Physics, Hokkaido University, Sapporo
060-0810, Japan}

\author{Y. Toda}
\affiliation{Department of Applied Physics, Hokkaido University,
Sapporo 060-0810, Japan} \affiliation{Division of Innovative
Research CRIS, Hokkaido University, Sapporo 001-0021, Japan}

\author{ K. Shimatake}
\affiliation{Department of Applied Physics, Hokkaido University,
Sapporo 060-0810, Japan} \affiliation{Division of Innovative
Research CRIS, Hokkaido University, Sapporo 001-0021, Japan}

\author{N. Momono}
\affiliation{Department of Materials Science and Engineering,
Muroran Institute of Technology, Muroran 050-8585, Japan}

\author{M. Oda}
\affiliation{Department of Physics, Hokkaido University, Sapporo
060-0810, Japan}
\author{M. Ido}
\affiliation{Department of Physics, Hokkaido University, Sapporo
060-0810, Japan}

\begin{abstract}
We report the ultra-fast optical response of quasi-particles (QPs)
in both the pseudogap (PG) and superconducting (SC) states of
underdoped (UD) Bi$_{2}$Sr$_{2}$CaCu$_{2}$O$_{8+\delta}$ (Bi2212)
single crystal measured with the time-resolved pump-probe technique.
At a probe energy $\hbar\omega_{pr}$=1.55 eV, it is found that the
reflectivity change $\Delta$R/R changes its sign at exactly $T_{c}$,
which allows the direct separation of the charge dynamics of PG and
SC QPs. Further systematic investigations indicate that the
transient signals associated with PG and SC QPs depend on the probe
beam energy and polarization. By tuning them below $T_{c}$ two
distinct components can be detected simultaneously, providing
evidence for the coexistence of PG and SC QPs.

\end{abstract}
\date{\today}

\maketitle

The relationship between the anomalous PG state and high-$T_{c}$
superconductivity is still an important open issue in condensed
matter physics. Two main scenarios have been put forward to help us
to understand this relationship well within a theoretical framework.
One proposes that the PG state is dominated by a hidden order, which
competes with high-$T_{c}$ superconductivity. The other emphasizes
that the PG state is a precursor of high-$T_{c}$ superconductivity.
In this picture, high-$T_{c}$ superconductivity originates from the
PG state where Cooper pairs are formed but lack long-range phase
coherence, which will be established below $T_{c}$. In early
research, angle-resolved photoemission spectroscopy (ARPES) and
electron tunneling spectroscopy revealed that the PG could smoothly
evolve into the SC gap as the temperature fell below $T_{c}$, while
electronic Raman scattering (ERS) data showed that nodal and
antinodal gaps coexisted below $T_{c}$ \cite{Opel,Oda1}. In fact,
this discrepancy can be understood well in terms of the scenario of
the ``Fermi-arc'' superconductivity \cite{Oda1}, which is supported
by additional high-resolution data recently obtained again by ERS
\cite{Tacon} and by ARPES \cite{Tanaka,Kondo,Lee}, revealing that
for UD cuprate superconductors below $T_{c}$ there are two energy
scales in the nodal and antinodal regions on the Fermi surface. Such
a high consistency of different experimental data validates the
contention that the SC gap located in the nodal region, called
``Fermi-arc'' superconductivity
\cite{Tanaka,Oda,Oda1,McElroy,Hashimoto}, is distinct from the PG
located in the antinodal region for UD cuprate superconductors.

Distinguishing the charge dynamics of QPs in the antinodal and nodal
regions will be helpful as regards revealing the mechanism
responsible for them. Time-resolved pump-probe optical spectroscopy
with femto-second resolution is a powerful tool for determining the
charge dynamics of carriers in superconductors \cite{Brorson}. In
such measurements, a pump pulse with a higher energy than the SC
energy gap breaks Cooper pairs into two electrons (or photoexcited
carriers) and excites them to a non-equilibrium high-energy state.
Subsequently the probe pulse detects, within the delay time between
the pump and probe pulses, the process by which photoexcited
carriers recombine into a superconducting condensate. Below $T_{c}$,
the PG does not change into the SC gap for UD cuprate
superconductors \cite{Opel,Oda1,Tacon,Tanaka,Kondo,Lee}, meaning
that the relaxation process of SC QPs (Cooper pairs) should differ
from that of PG QPs. Therefore, in principle this technique can be
used to distinguish different relaxation processes involving PG and
SC QPs in UD Bi2212 by measuring the sample reflectivity change
$\Delta$R/R. Previously pump-probe experiments revealed that
$\Delta$R/R could be either positive or negative
\cite{Han,Eesley1,Kabanov,Demsar,Murakami,Gedik}. In particular,
Mihailovic $et~al.$ \cite{Demsar,Stevens,Kabanov,Thomas} and
Murakami $et~al.$ \cite {Murakami} have already conducted a
self-consistent two-component analysis of the optical spectroscopy
data obtained by the pump-probe technique, demonstrating that PG QPs
coexist with SC QPs below $T_{c}$. However, a direct and unambiguous
identification of distinct charge dynamics associated with SC and PG
QPs still lacks. Recently, we have used two-color time-resolved
pump-probe optical spectroscopy to investigate NbSe$_{3}$ with two
charge-density wave (CDW) transitions occurring at different
temperatures \cite{Shimatake}, revealing that the two transitions
can be selectively detected at temperatures far below the CDW
transition temperatures. In this letter, we report the direct
identification of the charge dynamics of PG and SC QPs by measuring
$\Delta$R/R of the UD Bi2212, using the same technique as that
described in Ref. \cite{Shimatake}.

UD Bi2212 single crystals with $T_{c}$=76 K (hole concentration
0.11) are grown by the traveling solvent floating zone method using
an infrared image furnace. The time-resolved reflectivity change
$\Delta$R/R is measured by using the pump-probe technique on the
sample mounted in a cryostat whose temperature ranges from 4 to 300
K. Both the probe and pump beams are crossed and incident in the
c-axis direction on freshly cleaved surfaces. The laser power
induced heating effect has been accounted for by measuring the
temperature dependence of the $\Delta$R/R amplitude at different
given powers. The pump-probe experimental configurations have been
described in detail elsewhere \cite{Shimatake}.

Figure 1(a) shows the temperature evolution of $\Delta$R/R as a
function of delay time measured at a probe energy
$\hbar\omega_{pr}$=1.55 eV and a pump energy $\hbar\omega_{pu}$=1.07
eV over a wide temperature range from 15 to 280 K. This waterfall
plot shows that the sign changes of $\Delta$R/R occur just at the SC
transition temperature $T_{c}$ and the PG-opening temperature
$T^{*}$, respectively. It should be pointed out that the $T^{*}$
determined here is consistent with the result of the tunneling
spectroscopy \cite{Dipasupil}. The sign of $\Delta$R/R below $T_{c}$
and above $T^{*}$ is positive, whereas it is negative between
$T_{c}$ and $T^{*}$. Hereafter, we define the positive (negative)
signal as one with a positive (negative) sign. Using a
single-component exponential decay function $\Delta$R/R($T$,
t)=A($T$)exp(-t/$\tau$), where A($T$) is the amplitude of
$\Delta$R/R as a function of temperature ($T$) and $\tau$ is the
relaxation time of QPs, we can fit the data well [solid lines in
Figs. 1(b), (c) and (d)] and achieve the relaxation times for
different QPs, whose temperature dependence is shown in Fig. 2(a).
Considering the noticeable sign changes of $\Delta$R/R present at
$T_{\rm c}$ and $T^{*}$, we naturally assign the positive component
appearing below $T_{\rm c}$ with a slow decay of $\sim$2.5 ps to SC
QPs (Cooper pairs), which corresponds to the recombination time of
Cooper pairs, consistent with those obtained on
Y$_{1-x}$Ca$_{x}$Ba$_{2}$Cu$_{3}$O$_{7-\delta}$ single crystal
\cite{Demsar} and UD Bi2212 film \cite{Murakami} using two-component
analysis. Another component that appears above $T_{\rm c}$ and fades
out at $T^{*}$ is ascribed to PG QPs with a relaxation time of
$\sim$0.5 ps. The third component appearing above $T^{*}$ is a
step-function response with a relaxation time of $\sim$0.8 ps [Fig.
1(d)], which is a typically bolometric effect in the metal
\cite{Easley} and similar to that observed in
YBa$_{2}$Cu$_{3}$O$_{7-\delta}$ (Y123) at 300 K \cite{Albrecht}.
This observation is consistent with ARPES results showing that  UD
Bi2212 is a metal above $T^{*}$ \cite{Norman1}. In this letter, we
focus our attention on the two components present below $T^{*}$. The
temperature dependence of the amplitude of $\Delta$R/R is shown in
Fig. 2(b), which shows two features: (1) the amplitude of the
positive SC component decreases with increasing temperature and
becomes zero at $T_{\rm c}$; (2) the negative PG component starts to
appear at $T_{\rm c}$ and fades out around $T^{*}$.

\begin{figure}[t]
\begin{center}
\includegraphics[scale=.38]{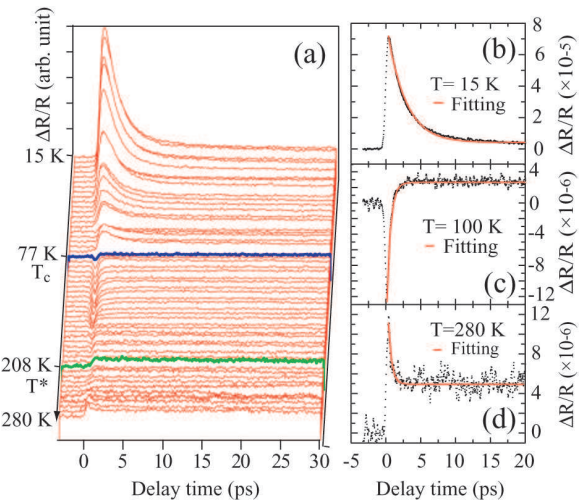}
\vspace*{-0.15cm} \caption{(Color online) (a) Waterfall plot for
$\Delta$R/R as a function of delay time over a wide temperature
range from 15 to 280 K measured at $\hbar\omega_{pu}$=1.07 eV and
$\hbar\omega_{pr}$=1.55 eV. The pump and probe beams are polarized
at 90$^{o}$ and 0$^{o}$ relative to the b axis, respectively. The
sign changes of $\Delta$R/R at $T_{c}$ and $T^{*}$ can be clearly
seen. Note that the temperature of each curve corresponds to that of
each point in Fig. 2. The single-component exponential function
fittings are for transient signals measured at 15 K (b), 100 K (c)
and 280 K (d). }\label{figure1}
\end{center}
\end{figure}

\begin{figure}[t]
\begin{center}
\includegraphics[scale=.36]{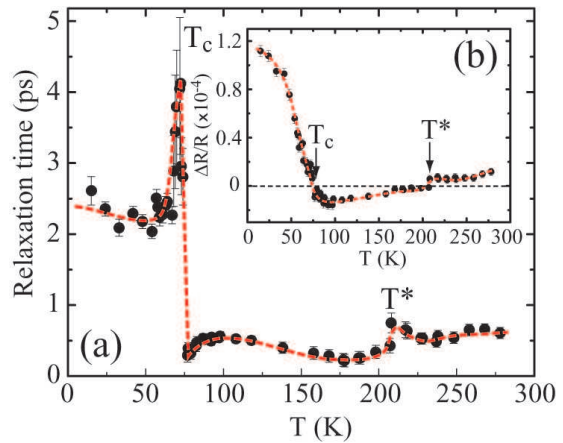}
\vspace*{-0.15cm} \caption{(Color online) Temperature dependence of
(a) the relaxation time of QPs, and (b) the measured amplitude of
$\Delta$R/R derived from the spectra shown in Fig. 1(a). The dashed
lines are guides to the eye.}\label{figure2}
\end{center}
\end{figure}

Figures 3(a) and (b) show the effect of the probe beam polarization
on $\Delta$R/R measured at $\hbar\omega_{pr}$=1.55 eV below and
above $T_{c}$. It is worthwhile noting that $\Delta$R/R shows no
dependence on the pump beam polarization, which is consistent with
previous reports on Y123 \cite{Stevens}, but is closely related to
that of the probe beam. Below $T_{\rm c}$ the transient $\Delta$R/R
is independent of the probe polarization and two identical signals
are obtained for \emph{${\textbf{E}_{pr}}$}$\parallel$ and
\emph{${\textbf{E}_{pr}}$}$\perp$
(\emph{${\textbf{E}_{pr}}$}$\parallel$ and
\emph{${\textbf{E}_{pr}}$}$\perp$ denoted as the probe polarizations
parallel and orthogonal to the b axis of the crystal,
respectively.). Above $T_{\rm c}$, however, the $\Delta$R/R measured
with \emph{${\textbf{E}_{pr}}$}$\perp$ is negative while that
measured with \emph{${\textbf{E}_{pr}}$}$\parallel$ is negligible,
which may be owing to the anisotropy of the probe transition matrix
elements \cite{Dvorsek}.

\begin{figure}[t]
\begin{center}
\vspace{0.2 cm}
\includegraphics[scale=.40]{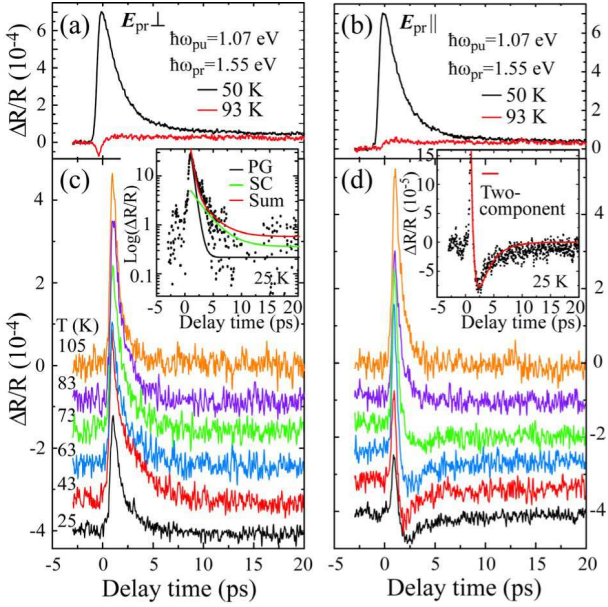}
\vspace*{-0.15cm} \caption{(Color online) The effect of probe
polarization on the signal. The transient $\Delta$R/R measured at
$\hbar\omega_{pu}$=1.07 eV and $\hbar\omega_{pr}$=1.55 eV both below
and above {\it T}$_{\rm c}$ with (a) \emph{$\textbf{E}_{pr}$}$\perp$
and (b) \emph{${\textbf{E}_{pr}}$}$\parallel$. The transient
$\Delta$R/R measured at $\hbar\omega_{pu}$=1.55 eV and
$\hbar\omega_{pr}$= 1.07 eV at various temperatures with (c)
\emph{${\textbf{E}_{pr}}$}$\perp$ and (d)
\emph{$\textbf{E}_{pr}$}$\parallel$. For clarity each spectrum
except for the topmost one is shifted by 8$\times$10$^{-5}$. The
inset in Fig. 3(c) shows the method used to analyze the signal
measured at 25 K. The black line corresponds to single PG-component
fitting, the green line to single SC-component fitting and the red
line to the sum of the above two fittings. The inset in Fig. 3(d)
shows the fitting using a two-component exponential decay function
for the signal measured at 25 K (red solid line). }\label{figure3}
\end{center}
\end{figure}

\begin{figure}[t]
\begin{center}
\includegraphics[scale=.35]{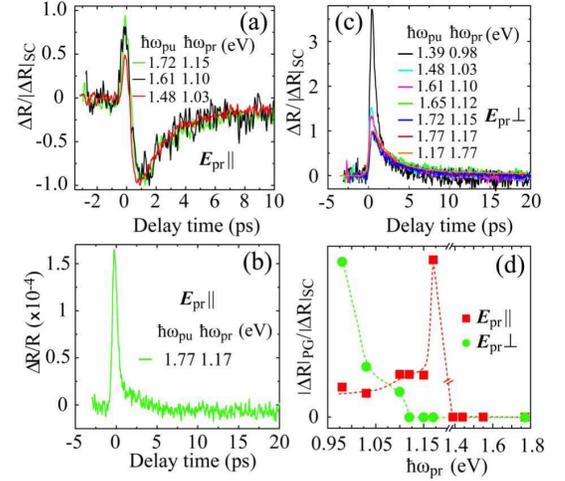}
\vspace*{-0.15cm} \caption{(Color online) The effect of probe energy
on the signal measured at 20 K. (a) Normalized $\Delta$R/R measured
at $\hbar\omega_{pr}$ from 1.03 to 1.15 eV with
\emph{${\textbf{E}_{pr}}$}$\parallel$. (b) Single PG component
measured at $\hbar\omega_{pr}$=1.17 eV with
\emph{${\textbf{E}_{pr}}$}$\parallel$. (c) Normalized $\Delta$R/R
measured $\hbar\omega_{pr}$ from 0.98 to 1.77 eV with
\emph{${\textbf{E}_{pr}}$}$\perp$. (d) Probe energy dependence of
the amplitude ratio of the PG and SC components. This ratio is
helpful in clarifying the change in the PG component with probe
energy. }\label{figure4}
\end{center}
\end{figure}

Further investigations indicate that the transient $\Delta$R/R also
strongly depends on probe energy $\hbar\omega_{pr}$. In Figs. 3(c)
and (d) are shown the data measured after exchanging the pump and
probe energies used above, i.e. measured at $\hbar\omega_{pu}$=1.55
eV and $\hbar\omega_{pr}$=1.07 eV. For
\emph{${\textbf{E}_{pr}}$}$\perp$ the sign of $\Delta$R/R measured
at $\hbar\omega_{pr}$=1.07 eV is positive throughout the whole
temperature range [Fig. 3(c)]. Above $T_{ c}$, the one-component
exponential decay function $\Delta$R/R($T$,
t)=A($T$)exp(-t/$\tau$$_{PG}$) is used to fit the data well,
resulting in the relaxation time $\tau$$_{PG}$$\sim$0.5 ps, which is
equal to that measured at $\hbar\omega_{pr}$=1.55 eV in the PG state
shown in Fig. 1(c). Since each kind of QP corresponds to one
relaxation time, it is natural to assign this component to PG QPs.
Below $T_ {c}$, the signal consists of two components that have slow
and fast decays, respectively. In this case, the analysis method is
similar to that used in Refs. \cite{Demsar,Murakami} and illustrated
in the inset of Fig. 3(c). We first use a single-component decay
function to fit the SC and PG components, respectively, by fixing
$\tau$$_{PG}$$=$0.5 ps and $\tau$$_{SC}$$=$2.5 ps and then obtain a
sum of the above two fittings. It can be seen that this method works
well and we therefore assign the fast-decay component to PG QPs and
the slow-decay component to SC QPs. For
\emph{${\textbf{E}_{pr}}$}$\parallel$ the detected signal consists
of two distinct components with opposite signs below $T_ {c}$, as
seen in Fig. 3(d); the fast-decay component is positive and the
slow-decay component is negative. A two-component decay function,
$\Delta$R/R($T$,
t)=A($T$)exp(-t/$\tau_{PG}$)-B($T$)exp(-t/$\tau_{SC}$), is used to
obtain a good fit with the data measured below {\it T}$_{\rm c}$,
using parameters of $\tau_{PG}$$\sim$0.5 ps and $\tau_{SC}$$\sim$2.5
ps. One fitting curve is shown as an example in the inset of Fig.
3(d). With increasing the temperature, the amplitude of the negative
component decreases and becomes zero at {\it T}$_{\rm c}$,
suggesting that it originates from SC QPs. Above {\it T}$_{\rm c}$
only the positive component remains whose relaxation time is about
0.5 ps, suggesting that the positive component observed below {\it
T}$_{\rm c}$ is due to PG QPs. Therefore, we directly observe the
coexistence of PG and SC QPs in UD Bi2212, which have different
relaxation dynamics. This time-domain observation of the coexistence
of PG and SC QPs below $T_{\rm c}$ is in good agreement with
real-space measurements such as those obtained with scanning
tunneling microscopy/spectroscopy (STM/STS) with atomic resolution,
which reveal that SC and PG gaps coexist in real space in UD Bi2212
\cite{McElroy,Hashimoto,Liu}. Here, we would like to emphasize that
the coexistence of PG and SC QPs does not mean the ``inhomogeneous
phase separation'' of the PG and SC states in real space. Recent
STM/STS results suggest that both the PG and SC QPs should be
distributed uniformly for a disorder-free UD Bi2212 sample below
$T_{\rm c}$ \cite{Hashimoto,Liu,Liu2}.

In order to obtain detailed information on the energy dispersion of
the charge dynamics of PG and SC QPs below $T_{\rm c}$, we further
investigate the effect of probe energy on the transient signal at 20
K, indicating that either the PG or SC component can be selectively
detected at a specified probe energy. Figure 4(a) shows the signals
measured at probe energies of 1.03 to 1.15 eV with
\emph{${\textbf{E}_{pr}}$}$\parallel$, which is normalized to the
amplitude of the SC component to minimize the effect of the pump
power on the signal. As is clearly seen, two components originating
with SC and PG QPs can be simultaneously detected, as discussed
above. More interestingly, with increasing $\hbar\omega_{pr}$ the
fraction of PG component increases. At $\hbar\omega_{pr}$=1.17 eV, a
single PG component is detected [Fig. 4(b)], i.e. the fraction of PG
component in the measured signal is almost {100\%}. As the probe
energy increases further, the PG-component fraction decreases and
the SC-component fraction increases and only a single SC component
is detected in the energy range from 1.39 to 1.77 eV. To see this
trend clearly, we show the amplitude ratio of the PG and SC
components $|$$\Delta$R$|$$_{PG}$/$|$$\Delta$R$|$$_{SC}$
($|$$\Delta$R$|$$_{SC}$ and $|$$\Delta$R$|$$_{PG}$ corresponds to
the amplitudes of the SC and PG components, respectively.) as a
function of $\hbar\omega_{pr}$ for
\emph{${\textbf{E}_{pr}}$}$\parallel$ by solid squares in Fig. 4(d).

The energy dependence of the signal measured with
\emph{${\textbf{E}_{pr}}$}$\perp$ differs from that measured with
\emph{${\textbf{E}_{pr}}$}$\parallel$. For
\emph{${\textbf{E}_{pr}}$}$\perp$ the signal is always positive
regardless of $\hbar\omega_{pr}$, as shown in Fig. 4(c). Careful
fitting analysis indicates that at $\hbar\omega_{pr}$=0.98 eV the PG
component dominates the signal. Increasing the probe energy results
in a decrease in the PG component fraction and an increase in the SC
component fraction; in the probe energy range from 1.12 to 1.77 eV,
only the SC signal is detected. This trend is illustrated in Fig.
4(d) by solid circles, where $|$$\Delta$R$|$$_{PG}$ and
$|$$\Delta$R$|$$_{SC}$ are obtained by fitting the data using the
method described in the inset of Fig. 3(c). Note that the signals
measured at $\hbar\omega_{pr}$=1.03 and 1.10 eV are similar to those
measured previously for La214 and Y123 samples \cite{Demsar,Rast}
that consist of the mixed SC and PG components.

The most important finding of our present investigation is the
direct observation of coexisting PG and SC QPs below $T_{c}$ [Fig.
3(d) and Fig. 4(b)], which agrees well with findings obtained using
electronic Raman scattering \cite{Opel,Oda1,Tacon}, ARPES
\cite{Tanaka,Kondo,Lee} and STM/STS \cite{McElroy,Hashimoto,Liu}
that two energy scales coexist below $T_{c}$. For UD Bi2212, the SC
gap located in nodal regions is distinct from the PG in antinodal
regions on the Fermi surface, which suggests that below $T_{c}$ the
PG does not evolve smoothly into the SC gap. As for the
time-resolved pump-probe measurement, to our knowledge, it is the
first time there has been an unambiguous and direct assignment of
the relaxation processes associated with PG and SC QPs based on the
sign changes of $\Delta$R/R at $T_{c}$ and {\it T}$^{*}$ [Fig.
1(a)], whose relaxation times are $\sim$0.5 ps and $\sim$2.5 ps,
respectively. At present, however, it is still difficult to infer
whether the PG state is a friend or foe of high $T_{c}$
superconductivity \cite{Norman}.

However, the present investigations clarify an important
long-standing puzzle with respect to the sign of the transient
$\Delta$R/R measured by time-resolved optical spectroscopy. We
verify that the sign of $\Delta$R/R depends both on
$\hbar\omega_{pr}$ and on the probe beam polarization , which may be
due to the anisotropy of the probe transition matrix elements and
the inter-band transition probability \cite{Dvorsek}. Both strongly
depend on the probe energy $\hbar\omega_{pr}$ and the band structure
of the material \cite{Dvorsek}. Therefore, a single SC or PG
component can be detected separately by tuning the probe
polarization and energy. This method for separating the
contributions of PG and SC QPs to $\Delta$R/R is useful for
investigating the charge dynamics of QPs in other high- $T_{c}$
superconductors in order to clarify the universal relationship
between the SC and PG states.

This work was supported by the 21st century COE program
``Topological Science and Technology'' and Grants-in-Aid for
Scientific Research from the Ministry of Education, Culture, Sports,
Science and Technology, Japan.

\end{document}